# IMPROVING POSITIONING ACCURACY IN WCDMA/FDD NETWORKS UTILIZING ADAPTIVE THRESHOLD FOR DIRECT COMPONENT DETECTION

Natasa Begovic[1] and Aleksandar Neskovic[2]

[1]Technology Department, Telenor, Belgrade, Serbia
[2]Department of Telecommunications
School of Electrical Engineering, Belgrade, Serbia

*ABSTRACT*

*In NLOS propagation conditions power of direct component can be attenuated significantly. Therefore detection of direct component is aggravated which can degrades accuracy of Time of Arrival mobile positioning. The goal of this paper is to determine possibilities to improve estimation of direct component time delay by reducing detection threshold. Three different methods for calculating threshold has been tested and compared in terms of positioning error.*

## 1. INTRODUCTION

Time of Arrival mobile positioning technique is based on the fact that distance between mobile and base station is proportional to time delay of signal traveling between mobile and base station. After obtaining minimum three estimated ranges (called pseudo-ranges) to non-collocated base stations, mobile position can be calculated as intersection point of three circles. Centres of the circles correspond to the coordinates of base stations and radii correspond to estimated ranges. Due to multipath propagation, received signal is a sum of numerous wave components coming at the receiver end, with different delays, levels and incidental angles. Distance between mobile and base station, has to be estimated using the time delay of direct wave component, i.e. component that travels along the straight path from the transmitting antenna to the receiving antenna. Time delay of direct component is minimal comparing to the delays of all other propagation components, while its power strongly depends on propagation conditions. If receiving power of the direct component is greater than the power contained in all the other components, propagation is probably, LOS (Line of Sight). In case of NLOS (Non LOS) propagation direct component is not power-dominant (in some situations can also be absent) on the receiving side. In WCDMA/FDD receiver, which employs RAKE, prior to combining energy of different multipath propagation components, channel impulse response has been estimated. It is clear that in case of LOS propagation, time delay associated with maximum power in estimated impulse channel response corresponds to the time delay of direct component, while direct component detection in NLOS conditions (which are more more often case, specially in urban propagation environment) is not an easy task. Positive time delay estimation bias introduced by wrong detection of direct component in NLOS propagation conditions is often marked as NLOS error. During the positioning procedure, at least 3 pseudo-ranges have to be estimated, one toward serving cell and two (at least) toward neighbouring cells. Unfortunately, it can not be expected that mobile terminal have LOS towards other base stations than its serving. Even then, LOS propagation toward serving base station in urban environment is not so probable. Finally, it can be concluded, that the main cause of positioning error using time of arrival based technique in NLOS propagation is aggravated detection of direct wave component.

In literature, different approaches are suggested on how to detect direct component from the estimated channel impulse response, in order to improve positioning accuracy. The most common way is to compare estimated power of channel components with calculated detection threshold. The first component with power above detection threshold is said to be the direct one. In [1] detection threshold is calculated as average value of channel estimation lags and direct component is detected as the first lag in estimated impulse channel response that fulfill detection criterion. In [2] direct component has been searched for in a lag window before the first RAKE finger. Practically, due to reduced RAKE receiver complexity, number of RAKE fingers is limited, so it captures only those paths bearing the highest power. Direct component doesn't have to be among those captured RAKE fingers, but it is reasonable to assume that it is situated before the first one. Observed taps of estimated channel impulse response (before the first RAKE finger), has been tested using Generalized Likelihood Ratio Test (GLRT) in order to discriminate between signal and noise. An expression for false alarm probability has been derived in terms of noise variance and detection threshold. Considering known noise variance, for a given value of false alarm probability detection threshold can be calculated. First sample with calculated/estimated GLRT greater than detection threshold is identified as direct component. It has to be said that method proposed in [2] does not take into account influence of noise variance estimation nor threshold for detecting first RAKE component, but adopts assumed values in analysis of probability of direct component detection.

The goal of this paper is to determine possibilities to improve estimation of direct component time of arrival by using adaptive threshold of RAKE receiver. Problem of time delay estimation of direct component is introduced in next section. Three different methods for calculating adaptive detection threshold are presented in 2.1, 2.2 and 2.3. Comparison of these methods is done by means of simulation, i.e. for each method for calculating detection threshold, positioning error is simulated, as well as it's statistical parameters. Simulation characteristics are given in the Section 3, and finally, results are presented in Section 4.

## **2. TIME DELAY ESTIMATION OF DIRECT COMPONENT**

Due to non-orthogonality of pseudorandom codes, multipath propagation, receiver noise, etc., channel impulse response estimated at the receiver contains not only propagation components, but noise as well. In order to discriminate between propagation and noise, RAKE receiver uses estimated impulse channel response components whose average power exceeds detection threshold. Therefore, if time delay of direct component is estimated as time delay that corresponds to maximum power tap in channel impulse response or even as a time of earliest RAKE finger, positive bias is introduced. Hence, estimated time delay is probably, greater than the real time delay of direct component. Consequently, determined range is probably, greater than the true distance between mobile and base station.

Approach suggested in this paper is to reduce detection threshold as much as possible and to adjust it to a value just above the noise level. Three different methods for calculating detection threshold have been investigated and compared in terms of positioning error statistics. It is worth to mention that observed thresholds are initially developed for standard RAKE purpose, i.e. for maximizing signal to noise ratio at the receiver output.

### 2.1. Method 1 detection threshold calculation [3]

According to pretty simple rule, detection threshold is $\Delta$ dB smaller than maximum in power delay profile. Value $\Delta=0$ represents range estimation using time delay that corresponds to propagation component with maximum power in estimated channel impulse response.

### 2.2. Method 2 detection threshold calculation [3]

Non-coherent averaging of power delay profile is taken place in the receiver in order to mitigate small-scale fading and it is given by:

$$z(n) = \frac{1}{K} \cdot \sum_{i=1}^{K} |h_i(n)|^2 \quad (1)$$

where $h_i(n)$ is complex amplitude of estimated impulse response in tap $n$, $K$ is total number of impulse responses to be averaged. Expected value $m_z(n)$ of random variable $z(n)$ is given by [4]:

$$m_z(n) = E\{z(n)\} = |h(n)|^2 + \sigma_h^2, \quad (2)$$

where $h(n)$ is average of estimated channel response and $\sigma_h$ is standard deviation of estimated channel response. Expression (2) implies that even if multipath component with time delay $n \cdot T_c$ is absent from the impulse channel response ($h(n) = 0$), $m_z(n)$ is different from zero. Therefore, positive bias introduced in formula 2 corresponds to the expected value of noise $m_z$, noise contained in $z(n)$:

$$m_{z,noise} = \sigma_h^2. \quad (3)$$

Mathematical theorem claims that sum of the squares of $2N$ independent real normally distributed random variables $w_i$ has $\chi^2$ distribution with $2N$ degrees of freedom [4]. If standard deviation of normal random variable is denoted by $\sigma_w$, standard deviation of $\chi^2$ variable $\sigma_\xi$ is given by:

$$\xi = \sum_{i=1}^{2N} w_i^2, \quad \sigma_\xi = 2\sqrt{N} \cdot \sigma_w^2 \quad (4)$$

As standard deviation of complex variable $h_i(n)$ is and it's imaginary and real part (with standard deviation $\sigma_h/\sqrt{2}$ each) are mutually independent, sum of $K$ complex variables in expression (1) can be replaced with sum of $2K$ real ones. When substitute $\sigma_h/\sqrt{2}$ for $\sigma_w$, $K$ for $N$ in (4) and divide by $K$ caused by averaging in (1), standard deviation of averaged power profile $z(n)$ can be calculated as:

$$\sigma_{z,noise} = \frac{1}{K} 2\sqrt{N} \cdot \frac{\sigma_h^2}{2} = \frac{\sigma_h^2}{\sqrt{K}} = \frac{m_{z,noise}}{\sqrt{K}} \quad (5)$$

Deviation of random variable $z(n)$ in (5) is a consequence of noise and actually, presents deviation of noise $\sigma_z$ contained in $z(n)$. Detection threshold $\theta$ proposed in [3] is calculated as linear combination of mean and standard deviation of noise contained in averaged received power profile $z(n)$. After substitution of previously determined statistics of noise in $z(n)$, detection threshold is:

$$\theta = a \cdot m_{z,noise} + b \cdot \sigma_{z,noise}$$
$$= m_{z,noise}\left(a + \frac{b}{\sqrt{K}}\right) \approx m_z\left(a + \frac{b}{\sqrt{K}}\right), \quad (6)$$

where *a* and *b* are linear coefficients. If the total number of multipath components is much larger than the number of multipath components, which correspond to true transmission paths, which is usually the case, $m_{z,noise}$ is approximated with $m_z$ in (6) ($m_{z,noise} \approx m_z$).

### 2.3. Method 3 detection threshold calculation [5]

In method 3, detection threshold is calculated indirectly, by using estimated noise contained in power delay profile. Supposing that channel impulse response consist of *L* propagation components, noise is what is left behind when *L* strongest taps has been extracted from the estimated channel impulse response. Expression for detection threshold calculation is given by equation (7), whereby $m_z$ and $\sigma_z$ represent mean and standard deviation of estimated noise, respectively, and $\gamma$ stands for linear scaling factor [5].

$$\theta = m_{z,noise} + \gamma \cdot \sigma_{z,noise} \quad (7)$$

## 3. SIMULATION

In order to analyse accuracy of mobile positioning which employs previous detection thresholds, simulation model in Microsoft Visual C++ and Matlab has been developed. In WCDMA/FDD network, TOA positioning technique is based on measuring the time during which signal travels from base station to mobile and back to the base station. This time is denoted as RTT (Round Trip Time) parameter [6]. Under ideal propagation conditions and without any delay experienced in mobile terminal, distance between mobile and base station is proportional to the half of RTT parameter. Assuming symmetry (which is not ideal, but nevertheless exists) in signal delay from base station toward mobile terminal and vice versa, simulation considered only direction from base station to mobile (downlink). Therefore, only half of RTT parameter has been estimated and it is assumed that mobile knows when base stations start transmitting. Simulation environment consisted of 19 omni base stations, transmiting their own pilot signals Base stations are positioned at the centers of imaginary hexagons at mutual distance of 1km. Mobile moved randomly with constant speed of 50km/h. Urban propagation conditions have been simulated since urban environment is the most critical one (considering NLOS positioning error). Simulation employed complex propagation channel model which included the following propagation effects: basic path loss, multipath propagation and LOS/NLOS transitions. Analytical expression for path loss over smooth, plane terrain, 2GHz frequency and omni antenna is given by [7]:

$$L_P = \beta + \alpha * 10 \log(d), \quad (8)$$

whereby *α* and *β* stand for parameters that depend on the type of propagation environment, and *d* is distance between transmitting and receiving antenna.

On the other hand, each multipath component is characterized with it's time of birth, lifespan and time of disappearance, as well as it's complex amplitude, time delay and azimuth. During the movement of the mobile, amplitude, time delay and azimuth of each propagation component are changing according to the geometrical approach given in [8]. Small-scale fading is modeled through sinusoidal time evolution of component amplitude, during the half period sine, which represents lifespan of certain component [9]. Average number of multipath components is 5 (LOS) and 25 (NLOS) [8]. In case of LOS propagation, simulated propagation channel constains direct component, with zero time dispersion. It was assumed that power of direct component is 6dB greater then power of all other existing components [10]. In NLOS conditions, power-dominant direct component is absent. Probability of LOS propagation conditions toward serving base station is 20% [11], while propagation toward neighboring base stations is always NLOS. Mobile terminal measures time delays of pilot signals coming from 7 nearest base stations. Signifficant effect that influence the mobile positioning is hearability of distant base stations if a mobile is close to it's serving base station. As hearability is not subject of this paper, in order to eliminate it's impact, simulation employed IPDL (Idle Perid on DownLink) [3].

Impulse channel response is estimated by crosscorrelating the received signal with known base stations scrambling code. Length of scrambling code that is used to feed the correlation is one time slot. During measurement period, which lasts 30 frame-s, power delay profiles (squared impulse responses), are non-coherently averaged. Time delay of direct component is estimated as the time delay of the first component in averaged power delay profile channel impulse respond that exceeds detection threshold. Distance $D_k$ between mobile and k-th base station can be determined according to folowing:

$$D_k = c \cdot \tau_k = \sqrt{(x - x_k)^2 + (y - y_k)^2} \quad k=1..7, \tag{9}$$

wherby $c$ stands for speed of light, $(x,y)$ for coordinates of mobile to be determined, $(x_k, y_k)$ for known coodinates of k-th base station, and $\tau_k$ for estimated time delay of pilot signal coming from k-th base station. System of linear equations (10) can be derived, by substracting squared equation (9) for k=1, from other squared equation (9) for k=2..7. Employed method for solving linear system is least square.

$$\begin{aligned}(x_k - x_1)x + (y_k - y_1)y = \\ \frac{1}{2}(x_k^2 + y_k^2 - x_1^2 - y_1^2 + D_1^2 - D_k^2)\end{aligned}, k=2,3..7 \tag{10}$$

However, as the number of equations in (10) is greater then the number of unknowns, system is predetermined. In order to optimize the mobile position estimation, residuals from [12] has been used. Accordingly, from total set of equations (10), all possible subsets containing 3, 4, 5, 6 equations are formed. Each subset $q$ has it's own solution $(\hat{x}_q, \hat{y}_q)$ as well as residual defined by (11), where $S_q$ stands for set of equations that belongs to subset $q$. Residual is defined as sum of squared differences between measured ranges $r_m$ and ranges estimated using calculated position $(\hat{x}_q, \hat{y}_q)$ [12].

$$\hat{R}(q) = \sum_{m \in Sq} \left[ r_m - \sqrt{(\hat{x}_q - x_m)^2 + (\hat{y}_q - y_m)^2} \right]^2 \tag{11}$$

Final mobile position is equal to the subset position $(\hat{x}_q, \hat{y}_q)$ with the smallest residual. Positioning error has been calculated in 1020 simulation points, that are uniformly distributed all over the serving cell.

International Journal of Wireless & Mobile Networks (IJWMN) Vol. 7, No. 6, December 2015

## 4. RESULTS

Simulation results are presented through basic statistical parameters: mean, standard deviation and 95 percentile of the positioning error. 95 percentile of positioning error for calculating detection threshold method 1, 2, and 3 of has been presented in figures 1, 2 and 3, respectivelly.

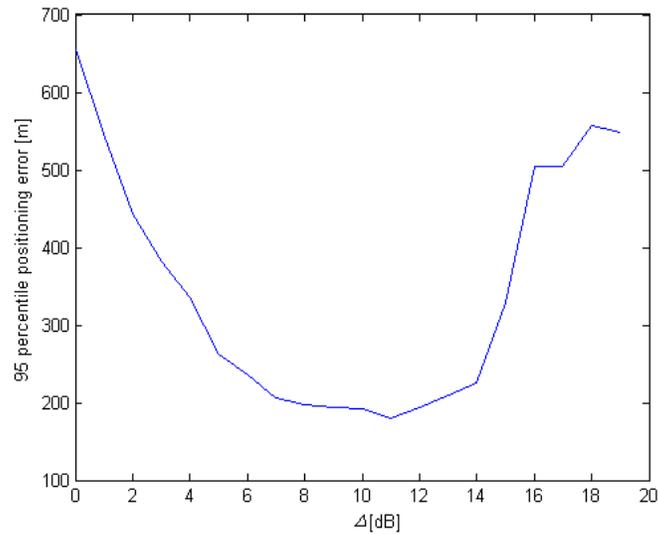

Figure 1. 95 percentile of positioning error as a function of parameter $\Delta$ (method 1)

With respect to calculating detection threshold method 1, for $\Delta<\Delta_{opt}$, value of estimated detection threshold is too high to detect strongly attanuated direct component, what causes positive bias in time delay estimation. On the other hand, in case $\Delta>\Delta_{opt}$ detection threshold is below noise level and estimated time delay is smaller than time delay of direct component. The minimal positioning error is achieved for $\Delta=11$.

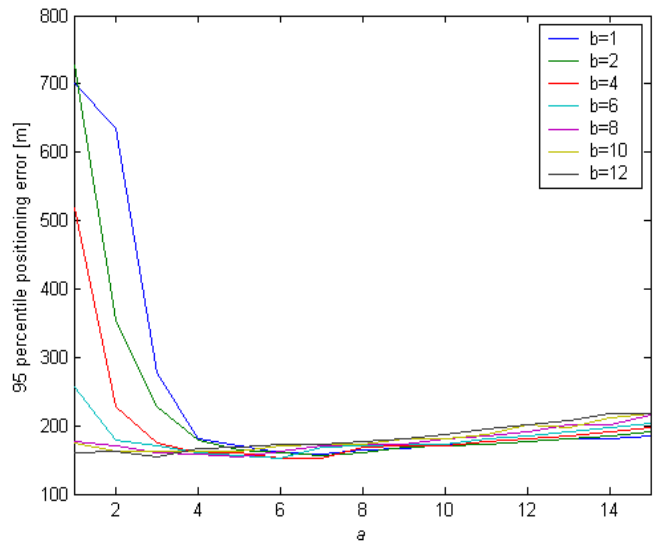

Figure 2. 95 percentile of positioning error as a function of parameters (a,b) (method 2)

It can be seen from the figure 2, that for calculation method 2, for each constant value of parameter *b*, increase of parameter *a*, causes curve to drop down, until optimal value, after which curve rises up, slowly. This results that for small values of parameters (*a*,*b*), estimated detection threshold is smaller than noise level, and hence estimated time delay is smaller than time delay of direct component. After reaching optimal positioning error, further increase of parameter *a*, will result in estimated time delay greater than the direct component, and smoother grow of 95 percentile of positioning error. The choice of parameters (*a*,*b*) equal to (6,4) secure best accuracy in terms of overall error statistics (mean, standard deviation and 95 percentile).

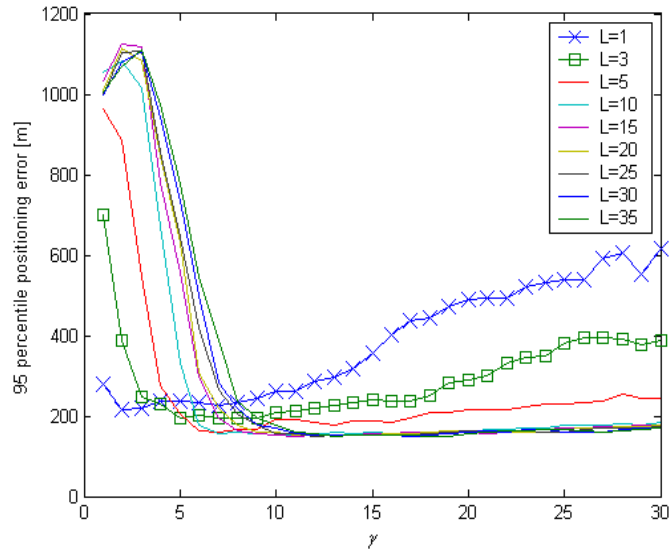

Figure 3. 95 percentile of positioning error as function of parameters *(L,$\gamma$)* (method 3)

Result for calculating detection threshold method 3, indicate that for a given *L*, 95 percentile of positioning error is greater for small values of parameter $\gamma$. For small values of parameter $\gamma$ and large values of *L*, detection threshold is pretty low, below the noise level, what causes negative error in time delay estimation of direct component. Increase of parameter $\gamma$ causes increase of detection treshold, reduction of error in time delay estimation and hence reduction in mobile positioning error. After reaching optimal result, further increase of $\gamma$ will effect on grow or stagnation of positioning error, depending on value of *L*. For *L*<10, increase of $\gamma$ implicate increase in detection treshold as well as increase of positive time delay estimation error and overall positioning error. On the other hand, for values *L*≥10, some components that are excluded from the power profile and treated as propagation components have power that is compared to the power of noise only. Therefore, increasing the number of propagation components will not significantly influence on value of calculated detection threshold. In this case, consequence of almost constant values of detection thresholds are constant values of estimated time delays and positioning error statistics. However, in practice, it can not be expected to have number of propagation components greater than 20 and hence setting of parameter *L* to a large value (*L*>20) is not reasonable. According to simulation conditions and obtained results average number of significant multipath component is up to 10, although the average number of components in simulated NLOS propagation channel is 25. During the simulation, it was seen that for $\gamma$>12 in negligible number of positions (<1.3‰), calculated detection thresholds were to high and it was not possible to detect, at least 3 direct wave components and to calculate mobile position.

According to previous, within each discussed method for calculating detection threshold, there are optimal parameter values that give the best statistics of positioning accuracy. Therefore, in order to compare the positioning accuracy of mentioned detection thresholds, the results assuming optimal parameter values are presented in table 1. Positioning error for trivial detection of direct component as maximum power component in channel response (Meth. 1 $\Delta=0$) is also given in table I. In compliance with table 1, all three detection threshold indicate better accuracy comparing to the trivial detection of direct component, as expected.

Table 1. Comparison of positioning error statistics for different methods of calculating detection thresholds

|  | *Meth. 1* | | *Meth. 2* | *Meth. 3* |
| --- | --- | --- | --- | --- |
|  | $\Delta=0$ | $\Delta_{opt}$ | $(a,b)_{opt}$ | $(L,\gamma)_{opt}$ |
| **Mean [m]** | 220 | 61 | 47 | 44 |
| **Std [m]** | 254 | 71 | 63 | 63 |
| **95 perc. [m]** | 655 | 197 | 153 | 150 |

## 5. SUMMARY AND CONCLUSIONS

This paper adresses improvement in estimation of direct component time of delay in NLOS propagation conditions, by using adaptive detection thresholds and hence better mobile positioning accuracy. It is showed that significant improvement can be achieved, by optimizing parameters within each discussed detection threshold. The best result, in terms od 95 percentile of positioning error, is achieved by using method 3 for calculating detection threshold, although method 2 has similar statistics. With respect to the computational complexity, method 1 is the simplest, method 2 follows and method 3 is the most complicated.

## ACKNOWLEDGEMENTS

Part of this work was carried out at the EURECOM (Sophia Antipolis, France), under the supervision of professor Dirk Slock.